\documentclass[apjl]{emulateapj}
\usepackage{epsf}
\usepackage{amssymb,amsmath}
\usepackage{multirow}
\usepackage{rotating}
\usepackage{subfigure}
\usepackage{color}

\def\aap{A\&A}
\def\apj{ApJ}

\def\apjl{ApJL}

\def\mnras{MNRAS}
\def\araa{ARA\&A}
\def\aj{AJ}

\def\pasj{PASJ}
\def\pasp{PASP}

\def\apss{Ap\&SS}

\begin{document}
\title{The Rate of Gas Accretion onto Black Holes Drives Jet Velocity}
\slugcomment{Accepted to ApJ Letters on 17 December 2014}
\author{Ashley~L.~King\altaffilmark{1,2}, 
Jon~M.~Miller\altaffilmark{3}, 
Michael Bietenholz\altaffilmark{4,5}, 
Kayhan G\"ultekin\altaffilmark{3}, 
Mark Reynolds\altaffilmark{3}, 
Amy Mioduszewski\altaffilmark{6}, 
Michael Rupen\altaffilmark{7}, 
Norbert Bartel\altaffilmark{4} }

\altaffiltext{1}{Department of Physics, 382 Via Pueblo Mall, Stanford, CA 94305, ashking@stanford.edu}
\altaffiltext{2}{Einstein Fellow}
\altaffiltext{3}{Department of Astronomy, University of Michigan, 1085 S. University Ave, Ann Arbor, MI 48109-1107, USA}
\altaffiltext{4}{ Department of Physics and Astronomy, York University, Toronto, M3J~1P3, Ontario, Canada}
\altaffiltext{5}{ Hartebeesthoek Radio Observatory, PO Box 443, Krugersdrop, 1740, South Africa}
\altaffiltext{6}{ National Radio Astronomical Observatory, P.O. Box O, Socorro, NM, 87801, USA }
\altaffiltext{7}{ NRC Dominion Radio Astrophysical Observatory, Penticton, British Columbia V2A 6J9}

\begin{abstract}
Accreting black holes are observed to launch relativistic, collimated jets of matter and radiation. In some sources, discrete ejections have been detected with highly relativistic velocities. These particular sources typically have very high mass accretion rates, while sources lower knot velocities are predominantly associated with black holes with relatively low mass accretion rates. We quantify this behavior by examining knot velocity with respect to X-ray luminosity, a proxy for mass accretion rate onto the black hole. We find a positive correlation between the mass-scaled X-ray luminosity and jet knot velocity. In addition, we find evidence that the jet velocity is also a function of polar angle, supporting the ``spine-sheath" model of jet production.  Our results reveal a fundamental aspect of how accretion shapes mechanical feedback from black holes into their host environments.
\end{abstract}

\maketitle


\section{Introduction}
Accreting black holes produce highly collimated, relativistic outflows known as jets. These jets inject enormous quantities of material and energy into their host galaxies \citep{Allen06}, creating shocks, heating gas, and ultimately transforming their galactic and extra-galactic environments \citep{Fabian12}. Though we can observe the impact of jets on their surroundings, we have only measured the total power from the most massive active galactic nuclei (AGN) jets via the X-ray and radio jet cavities they carve out \citep{Allen06}. Smaller AGN jets do not leave the same observable signatures as their high-mass counterparts, leaving our understanding of jet kinematics in typical AGN jets wanting. 

Besides measuring jet power from a select few excavated cavities, kinematic power can be estimated via observations of the moving knots within the jet itself. This has the added benefit of being an instantaneous power measurement rather then a time averaged estimate, and is not necessarily limited to the highest mass systems. Current observations of jet velocities qualitatively suggest an important dependence on mass accretion rate, i.e., jet production at low mass accretion rates is mildly-relativistic \citep{Ulvestad99}, while quasars and blazars -- supermassive black holes with high mass accretion rates, $L_X\sim L_{Edd}$ -- produce jets with discrete, highly relativistic knots \citep{Chatterjee09,Chatterjee11}. This suggests an integral relation between the black hole and its accretion disk to launching, collimating, and accelerating jets. Consequently, dichotomies in jet radiative power, kinetic power, and morphological structures would readily be explained via intrinsic properties of the jet rather than extrinsic properties of its environment \citep{Ghisellini01}.

A detailed study of an ensemble of jet kinematic properties as a function of accretion rate has yet to be undertaken. We have assembled a sample of sixteen sources in a redshift range of  $0.02<z<1.05$, with well determined (apparent) jet knot velocities, black hole masses, and X-ray luminosities, in order to test intrinsic jet characteristics as a function of disk accretion properties. 

\section{Sample Selection}
\subsection{X-ray selection}
We compiled a sample of sixteen AGN with discrete jet knot emission, black hole mass estimates, X-ray luminosity measurements between 2--10 keV, which were at a redshift of $0.02<z<1.05$. Table 1 lists the characteristics of our sample. The X-ray observations were taken in the  2--10 keV range in order to avoid absorption contamination by hydrogen gas along the line of sight, typically associated with lower energies. We find that rather than the X-ray luminosity, it is the X-ray luminosity scaled by each black hole's Eddington luminosity that is the relevant quantity in our analysis. A black hole's Eddington luminosity is a function of its mass, $L_{Edd}\simeq1.3\times10^{38} {M_{BH}}/{M_\odot}$.

A bolometric correction can translate the X-ray luminosity into the bolometric luminosity, $L_{Bol}=\kappa L_{X}$. The bolometric luminosity is the total accretion luminosity, which is a function of mass accretion rate ($L_{Bol}=\eta\dot{M}_{acc}c^2$, where $\eta$ is the radiative efficiency, $\dot{M}_{acc}$ is the mass accretion rate and $c$ is the speed of light). The bolometric correction may itself depend on and scale positively with bolometric Eddington fraction \citep{Vasudevan07}, and the efficiency is a monotonic function of the mass accretion rate scaled by mass \citep{Narayan98}. The efficiency also depends on the type of accretion flow in the accretion disk \citep{Narayan98}. In this study, we refrain from converting directly from X-ray Eddington fraction to mass accretion rate to avoid introducing larger uncertainties from both the bolometric and efficiency corrections. 

We also assumed the X-ray luminosity does not suffer from Doppler beaming from the jet, and is therefore independent of inclination effects. This is supported by the fact that we do not find a significant correlation between X-ray luminosity and inclination (see below). Finally, we used a 0.3 dex uncertainty for the X-ray luminosity to account for the variability inherent to AGN and uncertainties in the spectral shape. The masses also included an uncertainty of 0.3 dex, which not only accounts for intrinsic statistical errors but also systematic offsets between different mass estimate methods, i.e., power spectral density breaks \citep{Marscher04}, reverberation mapping \citep{Peterson04}, the $M_{BH}-\sigma$ relation \citep{Barth03}, Keplerian modeling of gaseous disk \citep{Ferrarese99}, and full-width half-maximum of emission lines \citep{Ghisellini10}.

\subsection{Jet Knot Velocities}
To compare with the X-ray Eddington fraction data, we collected the apparent velocities of discrete ejected knots, which are derived from Very Long Baseline Interferometric (VLBI) analyses, for these sources. A number of the sources have multiple knots with varying brightnesses and velocities. Previous studies have found that for those sources with at least ten detectable knots, each source has an apparent knot velocity distribution consistent with a Gaussian distribution and a mean velocity characteristic to each source \citep{Lister13}. In this study, we used only the brightest knot at $R\approx10^{5.5} R_G$ ($R_G=GM_{BH}/c^2$, where $G$ is the gravitational constant), the de-projected radius from the black hole. We find that the velocity of the brightest knot roughly corresponded to the mean velocity in those sources with multiple knots. 

\begin{figure*}
\centering
\hspace{0cm}
\subfigure[\label{Fig:rg}]{ \includegraphics[width=.5\linewidth,angle=0]{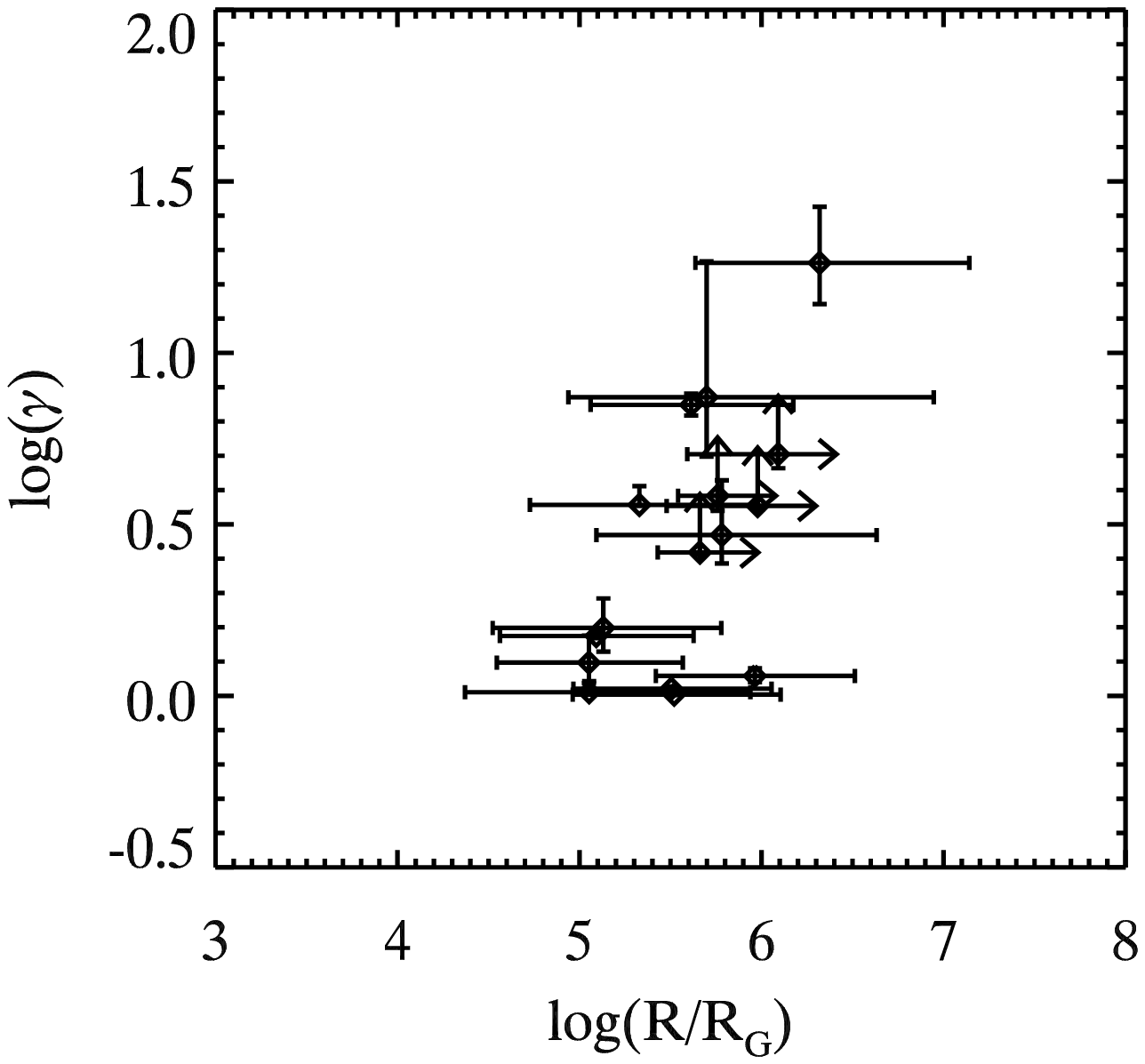}}
\hspace{-1cm}
\subfigure[\label{Fig:rgledd}]{\includegraphics[width=.5\linewidth,angle=0]{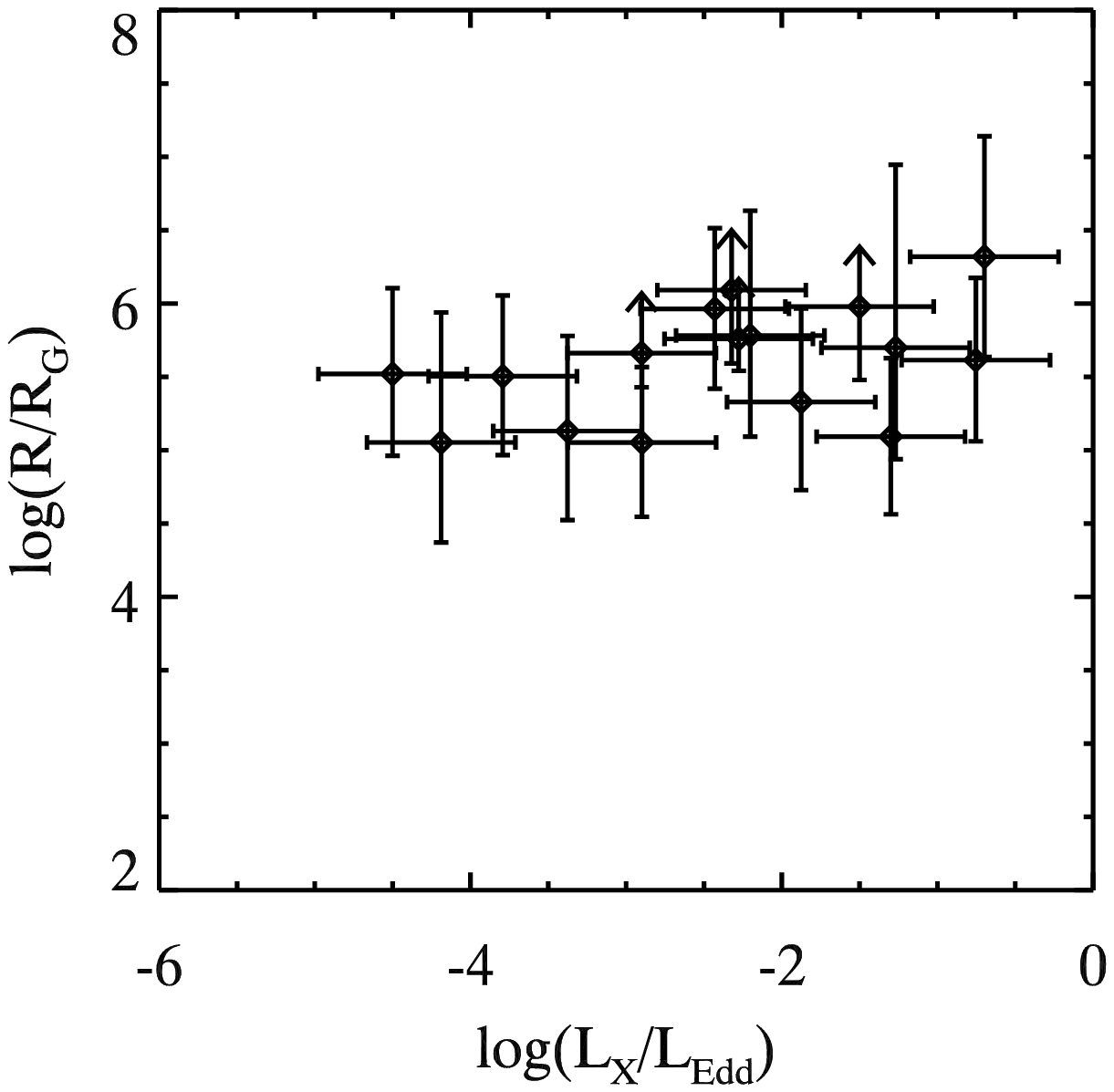}}
\caption{{\bf Jet Lengths} . {\bf a)} A logarithmic plot of the Lorentz factor plotted against the de-projected radius in terms of gravitational radii ($R_G$). {\bf b)} A logarithmic plot of the X-ray Eddington fraction versus the radius in terms of gravitational radii ($R_G$). There does not appear to be a trend  of X-ray Eddington fraction with radius. \label{fig:rg}}
\end{figure*}

Figures \ref{Fig:rg} \& \ref{Fig:rgledd} depict the radial distribution of these knots, and indicates no clear radial dependence in our data set. The restriction of the radius obe $R\simeq10^{5.5}$ R$_G$ served as the most stringent selection criterion. In addition, this procedure assumes that the brightest knot best characterizes the bulk motion in the jets, and it has the additional benefit of only sampling the jet close to the black hole. We note that if we instead use the maximum velocity knot instead of the brightest knot, the overall qualitative behavior of our relations still hold but with larger scatter.

The apparent velocities, $\beta_{app}$, are de-projected using the relation $\beta_{app} = \beta\sin\theta/(1-\beta\cos\theta)$, where $\theta$ is the inclination to our line-of-sight, and $\beta=v/c$ is the intrinsic velocity. The inclination of the jet to our line-of-sight is measured with $0^\circ$ as looking down the barrel of the jet and $90^\circ$ observing perpendicular to the jet flow. The inclinations were determined via a number of different methods. The most reliable method involved utilizing the ratio in either brightness or velocity between the approaching and receding knots \citep{Jones94,Giroletti06,Gentile07}. The angular separation between the approaching and receding knots was also used \citep{Asada06,Taylor09}. These methods assume that the jets are bi-polar outflows with simultaneous ejections at each pole. In addition, in some cases the upper limit on the inclination angle is set by the fact that the de-projected velocity cannot exceed the speed of light \citep{Gentile07,Chatterjee09}. Finally, a few sources had inclinations that were derived from averaging inclinations obtained from a number of knots, which were determined via the assumption that $\theta \lesssim \arcsin(1/\beta_{app)}$ \citep{Jorstad05}. We took care to ensure that regardless of the method, all the inclinations were determined in the vicinity of the knot measurements to minimize the effects of bending in the jets. Table 1 shows which methods were used on each source. 

Finally, we note that our sample contains varying jet morphologies, including both Fanaroff-Riley I (FR I) jets, which are jets dominated by compact, edge-brightened emission,  Fanaroff-Riley II (FR II) jets, which are edge-darkened and lobe-dominated jets \citep{Fanaroff74}, as well as BL Lacertae (BL LAC) objects, which are radio-loud jets which lack strong optical emission or absorption lines \citep{Urry95}.

\begin{figure*}
\subfigure[\label{Fig:lor}]{\includegraphics[width=.5\linewidth,angle=0]{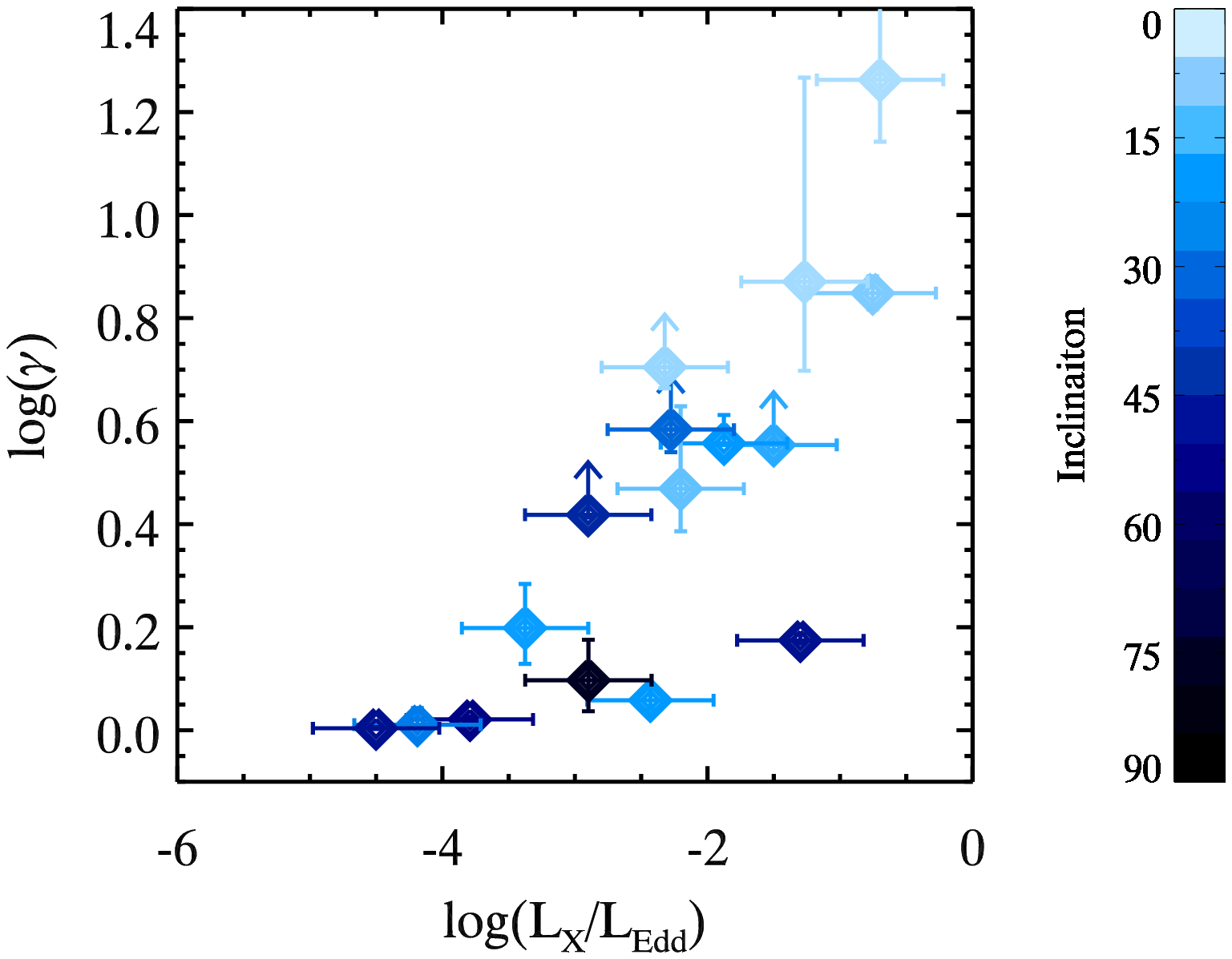}}
\hspace{-0.cm}
\subfigure[\label{Fig:inc}]{\includegraphics[width=.5\linewidth,angle=0]{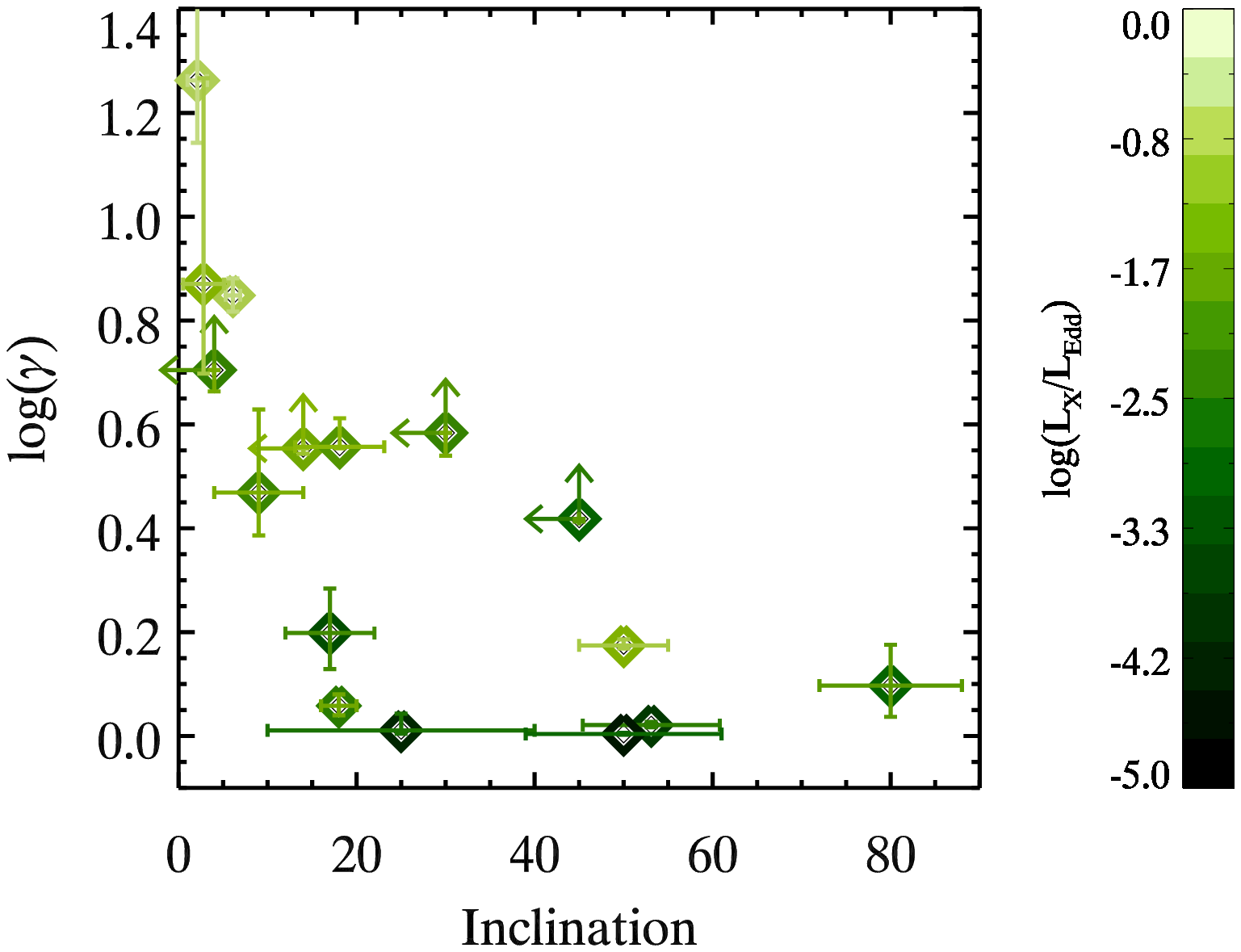}}

\caption{ {\bf Accretion driven jet velocity and structure.} {\bf a)} De-projected knot Lorentz factors ( $\gamma = (1-\beta^2)^{-1/2}$) of the brightest knots in AGN jets versus the X-ray Eddington fraction $L_X/L_{Edd}$), a proxy for mass accretion rate onto a black hole (see Sections 2.2 for jet knot velocity criteria). The blue color bar indicates the inclination angle to our line-of-sight of each source. There is a positive correlation between X-ray Eddington fraction and Lorentz factor (see Table 2). {\bf b)} Inclination versus Lorentz factor. The green color scale indicates $\log(L_X/L_{Edd})$. There is a negative correlation between inclination and Lorentz factor (see Table 2).
\label{Fig:vel} }
\end{figure*}

\section{Results}

In Figure \ref{Fig:vel}a, we plot the jet Lorentz factor, $\gamma = (1-\beta^2)^{-1/2}$, of the sample versus the X-ray Eddington fraction, $L_X/L_{Edd}$. The lower limits are a result of upper limits in inclinations, not an uncertainty in the velocity. We find that the Lorentz factor is positively correlated with the X-ray Eddington fraction, determined via a generalized Kendall's $\tau$ rank correlation test, which gives $\tau=0.55$ corresponding to 3.2$\sigma$ confidence level (see Table 2). This generalized Kendall's $\tau$ rank correlation test takes limits into account \citep{Isobe86}. When a partial correlation test is made to account for viewing angle as a potential third contributing factor (over plotted in blue in Figure \ref{Fig:vel}a), we find a 4.5$\sigma$ confidence level of correlation between X-ray Eddington fraction and Lorentz factor (see Table 2). In general, this indicates that as the X-ray Eddington fraction increases, faster knots are produced. 

We also plot the Lorentz factor against inclination in Figure 2b. There is a statistically significant, negative correlation between the two quantities, with a generalized Kendall's $\tau$ rank correlation test giving a $\tau=-0.47$ corresponding to a 3.5$\sigma$ statistical significance. Taking X-ray Eddington fraction into account as a third parameter (over plotted in green in Figure 2b), there is still a 3.3$\sigma$ confidence level of correlation between the viewing angle and the Lorentz factor that is observed.  An inverse correlation between inclination and Lorentz factor indicates that a faster velocity is observed when looking down the ``barrel'' of the jet as compared to viewing the jet from the side. This trend is to be expected if the jet structure is divided into a highly relativistic inner ``spine" and a mildly relativistic ``sheath", i.e., the ``spine-sheath'' jet model \citep{Attridge99}, as the Doppler boosting factor for the spine relative to that of the sheath is very large at low inclination, and very small at high inclinations \citep{Tavecchio08}.

We note that correlating the X-ray Eddington fraction with the viewing angle itself leads to only a small correlation. A generalized Kendall's $\tau$ test gives a $\tau=-0.36$ corresponding to a $2.2\sigma$ confidence level. This supports our use of X-ray Eddington fraction as a proxy for mass accretion rate, as it is not dominated by Doppler beaming from the jets.In addition, we also investigate whether the correlation itself is a result of any other hidden dependencies, specifically a distance dependence. The X-ray luminosity is dependent on the square of the luminosity distance, while the apparent velocity is linearly dependent on the luminosity distance. However, a partial correlation test between X-ray Eddington fraction and Lorentz factor gives a $3.3\sigma$ confidence level that correlation is not driven by distance to the source.

\section{Discussion}
The results have broad implications for jet structure and launching mechanisms. First, the results support the ``spine-sheath" structure model with a highly relativistic flow in the center of the jet surrounded by a mildly relativistic outer flow. This is evidenced by the fact that the de-projected velocity is partially dependent on viewing angle. Second, our analysis indicates that X-ray Eddington fraction also plays a prominent role in determining de-projected jet knot velocities. Assuming X-ray Eddington fraction is a monotonic function of mass accretion rate \citep{Narayan98} and bolometric corrections to the X-ray luminosities do not anti-correlate with jet velocity, this implies a coupling between material falling in toward the black hole and the energy that is extracted in the collimated jets. Generally, a higher influx of material results in faster knots ejected from the central AGN. 

One would expect such a relation if the mass accretion rate sets the magnetic field strength, $B$, and configuration, which in turn would set the bulk velocity. The former is likely to arise from equipartition of the accreted energy density with the magnetic field strength. The latter is predicted if the bulk velocity is set by the Alfv\'{e}n speed ($v^2_A\propto B^2/\rho$, where $\rho$ is the particle density) or magnetic hoop stresses \citep[e.g.,][]{Blandford82,Li02}. Future jet simulations are needed to explore this in more detail.

A range in velocities dependent on mass accretion rate, might partly explain the dichotomy between FR I and FR II jet structures. It has long been inferred that the difference between FR I and FR II jet structure is due to intrinsic differences in their jet velocities \citep{Ghisellini01}.  FR I jets would have slower velocities making them more susceptible to shearing instabilities, which result in a more diffuse, extended jet structure compared to the faster FR II jets. As FR I jets typically have lower Eddington fractions than FR II jets \citep{Ghisellini01}, our results therefore suggests that indeed FRI jets have lower velocities than FR II's,  further supporting this kinematically driven jet structure model. 

Looking across the black hole mass scale, a hint of Lorentz factor dependence on X-ray Eddington fraction has also been suggested for stellar-mass black holes \citep{Fender04}. This suggests a universal regulation mechanism for jet velocities that is set by X-ray mass accretion rate, as reflected by the X-ray Eddington fraction. However, as the stellar-mass work is based on velocity limits, more stringent observations are needed to determine whether the knot ejection velocities in stellar-mass black holes are actually dependent on X-ray Eddington fraction in the same manner as our supermassive black hole sample. 

A qualitatively similar trend between X-ray luminosity and jet power has also been discovered in the ``fundamental plane of black hole activity'', which finds a positive correlation between black hole mass, X-ray luminosity, and radio emission \citep[see][ and references therein]{Gultekin09}. Though radio emission is only a direct probe of the radiative power emitted by the jets, the fundamental plane has been used to infer a correlation between X-ray luminosity, mass, and total jet power (radiative plus kinetic). Our results directly measure a fundamental quantity contributing to the kinetic power, which dominates over the radiative power \citep{Allen06}. Assuming that the mass outflow rate ($\dot{M}$) is constant or increasing with X-ray Eddington fraction, then the kinetic jet power ($\gamma \dot{M}c^2$ ) also increases with X-ray Eddington fraction. However, the fundamental plane of black hole activity does indicate that radio luminosity, $L_R$, has a complex dependence on both X-ray luminosity and black hole mass, i.e., $L_R\propto L_X^{\alpha}M_{BH}^{\beta}$ \cite{Gultekin09}. Our sample has a relatively small mass distribution with an average mass of $\langle\log M_{BH}\rangle=8.77 M_\odot$ and a standard deviation of $\sigma=0.46$. Therefore, future investigations with higher resolution will need to expand our sample to even lower masses with the same extent as measured in gravitational radii, probing whether the jet velocities follow a similar dependence on both X-ray luminosity and mass. 

Finally, our results have interesting implications for jet launching mechanisms. Jet simulations suggest that a thick disk is needed to generate and collimate a relativistic jet \citep{Reynolds06,Mckinney12}. In the standard accretion theory paradigm, a thick disk only exists at low mass accretion rates (corresponding to $\lesssim 10^{-3}$L$_{\rm Edd}$), while a thin accretion disk exists at high mass accretion rates ($\gtrsim 10^{-3}$L$_{\rm Edd}$) \citep{Abramowicz13}. Our analysis shows coupling to the X-ray Eddington fraction up to very high X-ray Eddington fractions, where a thin disk should prevail. This indicates that either jets are just as efficiently coupled to thin disks as thick disks, or that thick disks prevail even at high X-ray Eddington fractions. Future investigations will need to further examine disk-jet theory in this context.

\acknowledgements We thank the anonymous referee for valuable comments. This research has made use of data from the MOJAVE database that is maintained by the MOJAVE team \citep{Lister13}. ALK would like to thank the support for this work, which was provided by NASA through Einstein Postdoctoral Fellowship grant number PF4-150125 awarded by the Chandra X-ray Center, which is operated by the Smithsonian Astrophysical Observatory for NASA under contract NAS8-03060


\begin{deluxetable*}{l l l l l l l l l l l }[h]
\tablecolumns{10}
\tablewidth{0pc}
 \label{tab:agn}
\tabletypesize{\scriptsize}
\tablecaption{Jet Knot Sample}
\tablehead{   Name  & $\beta_{app}$ & $\log R$ &$\theta$ & Method & $\log M_{BH}$& $\log(L_{X})$ & Type & z &  Ref.s\\
& & (pc) & $(^\circ$)&  & ($M_\odot$) & (erg s$^{-1}$) & (FR) & &}
\startdata
3C 48 &  3.70$\pm$0.40 &   1.0 &  $<$30 & DR,Max&  9.2 &  45.0 & I & 0.367 & $^{\bf 1,2}$ \\
3C 66A &  3.10$\pm$0.50&  0.2 &   $<$4.0& Max &   8.6 &  44.4 & BL & 0.444& 3,4,5 \\
3C 84 &  0.30$\pm$0.02 &   0.7 &  53.1$\pm$  7.7 & $v_{sep}$ & 8.6 &  42.9 & I & 0.0176 & 5,6,7\\
3C 111 &  3.44$\pm$0.08 &  -0.2 &  18.1$\pm$  5.0 & $\langle \theta\rangle$ & 8.3  &  44.5 & II & 0.0485 & 8,9 \\
3C 120 &  3.40$\pm$0.10 &  -0.2 &  $<14$ & Max &  7.7  &  44.3 & I & 0.033 & 9,10,11 \\
3C 273 &  6.70$\pm$0.39 &  0.0 &   6.1$\pm$  0.8 &  $\langle \theta\rangle$ & 8.7 &  46.0 & II & 0.158 & 5, 9,12 \\
3C 279 & 16.90$\pm$3.50 &   0.5 &   2.1$\pm$  1.1 & $\langle \theta\rangle$ &  8.9 &  46.3 & II & 0.536 & 2,5,13  \\
3C 338 &  0.11$\pm$0.01 &   1.1 &  50.0$\pm$ 11.0 & DR & 9.0 &  42.6 & I & 0.0304 & 14,15 \\
3C 390.3 &  2.42$\pm$0.03 &   0.9 &  $<$45 & Max &  9.0 &  44.2 & II & 0.0555& 5,16,17 \\
BL LAC &  2.07$\pm$0.16 &   0.3 &   9.0$\pm$  5.0 & DR &  8.6 &  44.5 & BL & 0.0686 & 5,18,19 \\
Cyg A &  0.66$\pm$0.26 &   1.1 &  80.0$\pm$  8.0 & DR & 9.4  &  44.6 & II & 0.0562 & 20,21,22  \\
Mrk 421 &  0.28$\pm$0.05 &   0.4 &  18.0$\pm$  2.0 & DR &  8.2 &  44.2 & BL & 0.0308 & 18,23,24 \\
Mrk 501 &  0.87$\pm$0.21 &   0.5 &  17.0$\pm$  5.0 & DR & 9.2 &  43.9 & BL & 0.0337 & 18,25 \\
NGC 6251 &  0.12$\pm$0.01 &  0.1 &  25.0$\pm$ 15.0 &  DR &  8.8 &  42.7 & I & 0.024  & 26,27,28,29\\
PKS 1741$-$038 &  4.70$\pm$1.10 &   0.6 &   2.8$\pm$  2.3 & A &  9.6 &  46.4 & II & 1.05 & 30,31 \\
PKS 1946+708 &  1.09$\pm$0.01 &   0.4 &  50.0$\pm$  5.0 & DR,$v_{sep}$ &   8.7 &  45.5 & II & 0.1008 & 32,33 \\
\enddata  

\tablecomments{\small{These are the brightest radio knots from each AGN  consistent with $\log R = 5.5 R_G$. We took the relative uncertainty in radius to be 0.2 dex, and an uncertainty of 0.3 dex in X-ray luminosity  and mass. The inclinations were determined via the following methods: DR: using Doppler ratio, $v_{sep}$: using opposing velocities, $A$: using angular separation of opposing knots to source, Max: requiring the de-projected velocity of the knot to be less than the speed of light, and $\langle \theta\rangle$: using the average angle of knots assuming $\theta\lesssim\sin^{-1}(1/\beta_{app})$. References: 1  \cite{An10}, 2 \cite{Wu13}, 3 \cite{Ghisellini10}, 4 \cite{Maraschi83}, 5 \cite{Lister13}, 6  \cite{Donato04}, 7 \cite{Asada06}, 8 \cite{Chatterjee11}, 9 \cite{Jorstad05}, 10 \cite{Peterson04}, 11 \cite{Chatterjee09}, 12 \cite{Marscher04}, 13 \cite{Fang05}, 14 \cite{Donato04},15 \cite{Gentile07}, 16 \cite{Gliozzi03}, 17 \cite{Dietrich12}, 18  \cite{Giroletti06}, 19 \cite{Ghisellini11}, 20 \cite{Young02}, 21 \cite{Tadhunter03}, 22 \cite{Bach05}, 23 \cite{Brinkmann05}, 24 \cite{Chen14}, 25 \cite{Barth03}, 26  \cite{Ferrarese99}, 27 \cite{Evans05}, 28 \cite{Migilori11}, 29 \cite{Jones94}, 30 \cite{Bassani13}, 31 \cite{Caproni14}, 32 \cite{Risaliti03}, 33 \cite{Taylor09}  }}

\end{deluxetable*} 

\begin{deluxetable}{l l l  }[h]
\tablecolumns{3}
\tablewidth{0pc}
\tabletypesize{\scriptsize}
\tablecaption{Rank Correlation Coefficients}
\tablehead{ Fit Parameters & Correlation & Confidence \\
& Coefficient & Level  }
\startdata

Kendall's $\tau$: & $\tau$ & $\sigma$ \\
$\log(\frac{L_{X}}{L_{Edd}})$, $\log\gamma$ & 0.55 & 3.2 \\
$\theta$, $\log\gamma$ & -0.47 & 3.5 \\ \hline

Partial Kendall's  $\tau$:  & $\tau$& $\sigma$ \\
$\log(\frac{L_X}{L_{Edd}})$, $\log\gamma$, {\bf ($\theta$) }& 0.45 & 4.5  \\  
 $\theta$, $\log\gamma$, {\bf ($\log(\frac{L_X}{L_{Edd}})$)} & -0.46 & 3.3 \\
$\log(\frac{L_X}{L_{Edd}})$, $\log\gamma$, {\bf (z) } & 0.46 & 3.3  
 \label{tab:agn}
\tablecomments{\small{ The table lists the rank correlation fits utilizing the Kendall's rank correlation test as well as a partial Kendall's correlation test. A positive $\tau$ indicates a positive correlation between the two variables, while a negative value indicates an inverse correlation. The significance of each correlation is also listed next to the magnitude of each correlation. The result of these correlation tests indicates that both X-ray Eddington fraction and inclination are integral in determining the inferred intrinsic velocity of the jet.  \label{tab:cor}}}

\end{deluxetable} 


\end{document}